\documentclass[11pt
,secnumarabic%
,tightenlines,amssymb, amsmath,nobibnotes,nofootinbib, aps, prl]{revtex4}
\usepackage[dvips]{graphicx}
\usepackage{tabularx}
\usepackage{bm}%
\expandafter\ifx\csname package@font\endcsname\relax\else
 \expandafter\expandafter
 \expandafter\usepackage
 \expandafter\expandafter
 \expandafter{\csname package@font\endcsname}%
\fi

\begin{document}
\title{Topological Phase Transitions and Holonomies in the Dimer Model}

\author{Charles Nash}
\affiliation{Department of Mathematical Physics, NUIM, Maynooth, Kildare, Ireland.}
\author{Denjoe O'Connor}
\affiliation{School of Theoretical Physics,
DIAS, 10 Burlington Road, Dublin 4, Ireland. }

\begin{abstract}
We demonstrate that the classical dimer model defined on a toroidal
hexagonal lattice acquires holonomy phases in the thermodynamic limit.
When all activities are equal the lattice sizes must be considered
${\rm mod}\ 6$ in which case the finite size corrections to the bulk
partition function correspond to a massless Dirac Fermion in the
presence of a flat connection with nontrivial holonomy.  For general
bond activities we find that the phase transition in this model is a
topological one, where the torus degenerates and its modular parameter 
becomes real at the critical temperature. We argue that these features are generic
to bipartite dimer models and we present a more general lattice whose
continuum partition function is that of a massive Dirac Fermion.
\end{abstract}

\maketitle

The classical statistical mechanics of lattice dimer models that interact
only via hard-core exclusion are among the  
simplest exactly solvable two dimensional models.
The subject has a long history and attracted the interest of both 
mathematicians \cite{Kenyon:2003uj} and physicists \cite{Nagle:1989}.
Recently, there have been several surprising advances and
these models have reemerged in new and diverse areas ranging from 
quantum dimer models \cite{Dijkgraaf:2008} 
to the melting corner/quiver gauge theory 
circle of ideas in string theory \cite{Okounkov:2003sp}. Furthermore, the dimer model on the hexagonal
lattice is closely related to the physics of graphene \cite{Novoselov:2005}.

Dimer models divide into two classes: Those defined on bipartite 
and non-bipartite lattices---a lattice is bipartite if its sites can be coloured black and white 
so that adjacent sites are always of opposite colour. The most 
general model then assigns a positive number, called an activity, 
to each bond.

In this note we focus on the simplest case of {\it dimers on
the hexagonal lattice}, with three distinct activities $a$, $b$ and $c$.
We present two principal observations:
\begin{itemize}
\item 
Surprisingly when all activities
are set to one the model has a mod 6 structure and the
limiting partition function is that of a free massless Dirac Fermion coupled to
flat connections. 
\item 
As the activities are varied there is a phase transition \cite{Nagle:1989}. 
We show that at this transition the geometry collapses in a topological phase transition. 
\end{itemize}
The model is critical when none of the activities
is larger than the sum of the other two.  
When considered on a torus, the finite size corrections 
are that of a free massless Dirac Fermion coupled to a flat connection. 
The angle between
the periods of the torus goes smoothly to zero so that the torus collapses to a degenerate 
one dimensional structure at the transition. This geometrical collapse
is surprisingly similar to that found recently in a very 
different context \cite{DelgadilloBlando:2007vx}.

It has been found \cite{Kenyon:2003ui,Kenyon:2003uj} that the phase
diagram of a bipartite dimer model is described by an {\it amoeba}
obtained from the zero locus of a certain spectral polynomial,
$p(z,w)$, that arises in the solution of the model on a torus as
described below.  The amoeba constitutes the parameter domain where
the model is critical.  In general it has compact and bounding phase
boundaries called ovals.

Our results imply that the general situation for bipartite dimer
models is therefore as follows: On the amoeba the partition function
has finite size corrections corresponding to that of a single massless
continuum Dirac Fermion on a torus, coupled to a flat connection.  The
dimer activities and periodic repetitions of the fundamental cell
determine the torus modular parameter and the holonomies of the flat
connection.  As the boundaries of the compact ovals are crossed there
is a phase transition in which the Fermion acquires a Dirac mass
\cite{Nash_and_OConnor:future}.  

The torus modular parameter, $\tau$, gradually changes as one
traverses the amoeba until finally as the outer boundary is crossed
there is another phase transition where the continuum geometry itself
collapses. This transition is continuous---there is no latent
heat---but asymmetric. As the transition is approached from the high
temperature side the specific heat diverges with critical exponent
$\alpha=1/2$, but there is no divergence as the transition is
approached from the low temperature side.

This transition is very similar to that discussed recently in a three
matrix model \cite{DelgadilloBlando:2007vx}. There, as the critical
temperature is approached from below, the spherical background geometry
evaporates with a diverging specific heat and exponent $\alpha=1/2$;
yet there is no divergence if the transition is approached from the
high temperature non-geometrical side.

\begin{figure}[hb]
 \includegraphics[height=30mm]{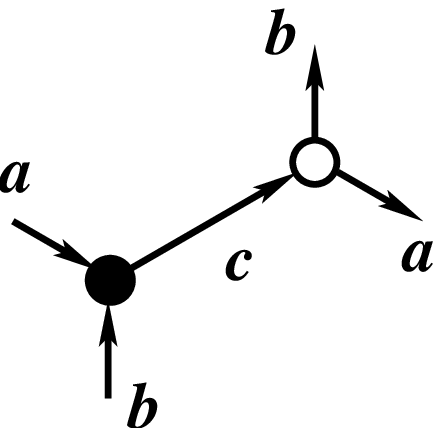}\qquad\qquad\qquad\qquad 
 \includegraphics[height=30mm]{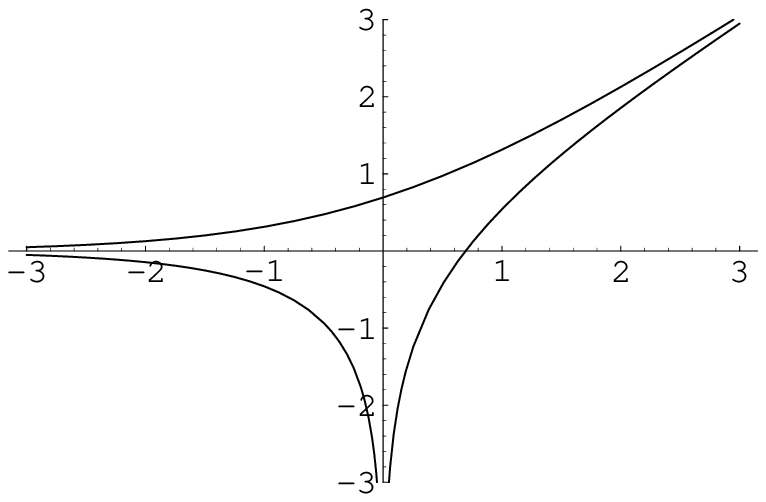}
 \caption{\footnotesize\label{fig:hex_simple} The basic tile for the hexagonal lattice, 
showing the activities and our choice of Kasteleyn orientation and its amoeba.}
 \end{figure}

We begin by constructing 
the Kasteleyn matrix for the $N\times M$ covering of a toroidal 
hexagonal lattice with fundamental tile given in Fig. \ref{fig:hex_simple}.
We then review the properties of the bulk partition function and discuss
the finite size corrections. The note ends with some speculations on
the six-vertex model.

If activities $a={\rm e}^{-\beta \epsilon_a}$, $b={\rm e}^{-\beta \epsilon_b}$ and 
$c={\rm e}^{-\beta \epsilon_c}$ are assigned to the bonds (c.f. Fig \ref{fig:hex_simple}) 
then the partition function of the model is given by
\begin{equation}
Z(N,M,a,b,c)=\sum_{coverings}a^{N_a}b^{N_b}c^{N_c}
\label{hafnian}
\end{equation}
where $N_i$ is the number of active bonds of type $i$ and $N_a+N_b+N_c=NM$, since
the lattice must be completely covered. 
When the activities are set to one, eqn. (\ref{hafnian}) counts the number of 
lozenge tilings of the dual triangular lattice. 

By assigning signs judiciously to the adjacency matrix,
Kasteleyn \cite{Kasteleyn:1961,Kasteleyn:1963}, Fisher and
Temperley \cite{Temperley:1961,Fisher:1961} observed that, one can convert 
(\ref{hafnian}) into a Pfaffian of, what is now called, a Kasteleyn matrix.  

The ``clockwise odd rule'' \cite{Fisher:1966} is used: Arrows are
placed on the links in such a way that by assigning $+1$ when
following an arrow and $-1$ when opposing one, the product of the
signs associated with any fundamental plaquette is $-1$ when the
plaquette is circulated in a counter clockwise direction.

The Kasteleyn matrix element $K_{AB}$ is then given by the activity
assigned to the link from vertex $A$ to vertex $B$ times $+1$ if the
arrow is from $A$ to $B$ but times $-1$ if it is from $B$ to $A$.

A Kasteleyn matrix, $K$, is therefore a signed weighted adjacency matrix 
for the lattice.  On any simply connected planar domain, the modulus of the
Pfaffian of $K$ gives the partition function of the dimer
model.

On a toroidal lattice the partition
function is given by a sum over the different discrete spin structures
and therefore involves four terms.  
This Kasteleyn matrix can in fact 
be viewed as a lattice Dirac operator \cite{Dijkgraaf:2007yr}, or more precisely a lattice
version of $C {\cal D}$ where $C$ is the charge conjugation matrix and
${\cal D}$ is the Dirac operator. In general a Kasteleyn matrix will describe
many lattice Fermions with different masses and the theory of 
classical dimer models is essentially equivalent 
to that of two dimensional lattice Dirac operators. 

A consistent sign assignment for the hexagonal lattice is for all arrows to come out of 
the black vertices and into the white vertices. 
However, this assignment will not give the standard association 
of spin structures to the resulting Kasteleyn matrix. So we choose a slightly different
convention with $a$ and $b$ bonds ingoing to the black vertex and the $c$ bond outgoing.
Then, for this hexagonal lattice, our choice of Kasteleyn matrix is
\begin{equation}
K=\left(\begin{array}{cc}
                  0&A \cr 
                   - A^T&0\cr
            \end{array}\right)
\end{equation}
where the matrix $A$ becomes
\begin{equation}
A_{k_1,k_2;l_1,l_2}=c\,\delta_{k_1,k_2}\delta_{l_1,l_2}-a\, \delta_{k_1,k_2+1}\delta_{l_1,l_2}-b\,\delta_{k_1,k_2}\delta_{l_1,l_2+1}
\end{equation}
This matrix can be diagonalized using the eigenvectors $z^k w^{-l}$ and one
finds 
\begin{equation}
A_{k_1,k_2;l_1,l_2} z^{k_2}w^{-l_2}=(c-a/z-b w)z^{k_1}w^{-l_1}
\label{spectral_A}
\end{equation}
so the eigenvalues are given by the spectral polynomial 
\begin{equation}
p^{Hex}(z,w)=c-a/z-bw \ .
\end{equation}
Curiously, this same polynomial also determines the phase diagram 
in Kitaev's quantum honeycomb model \cite{kitaev:2005}.

\begin{figure}[h]
 \includegraphics[height=30mm]{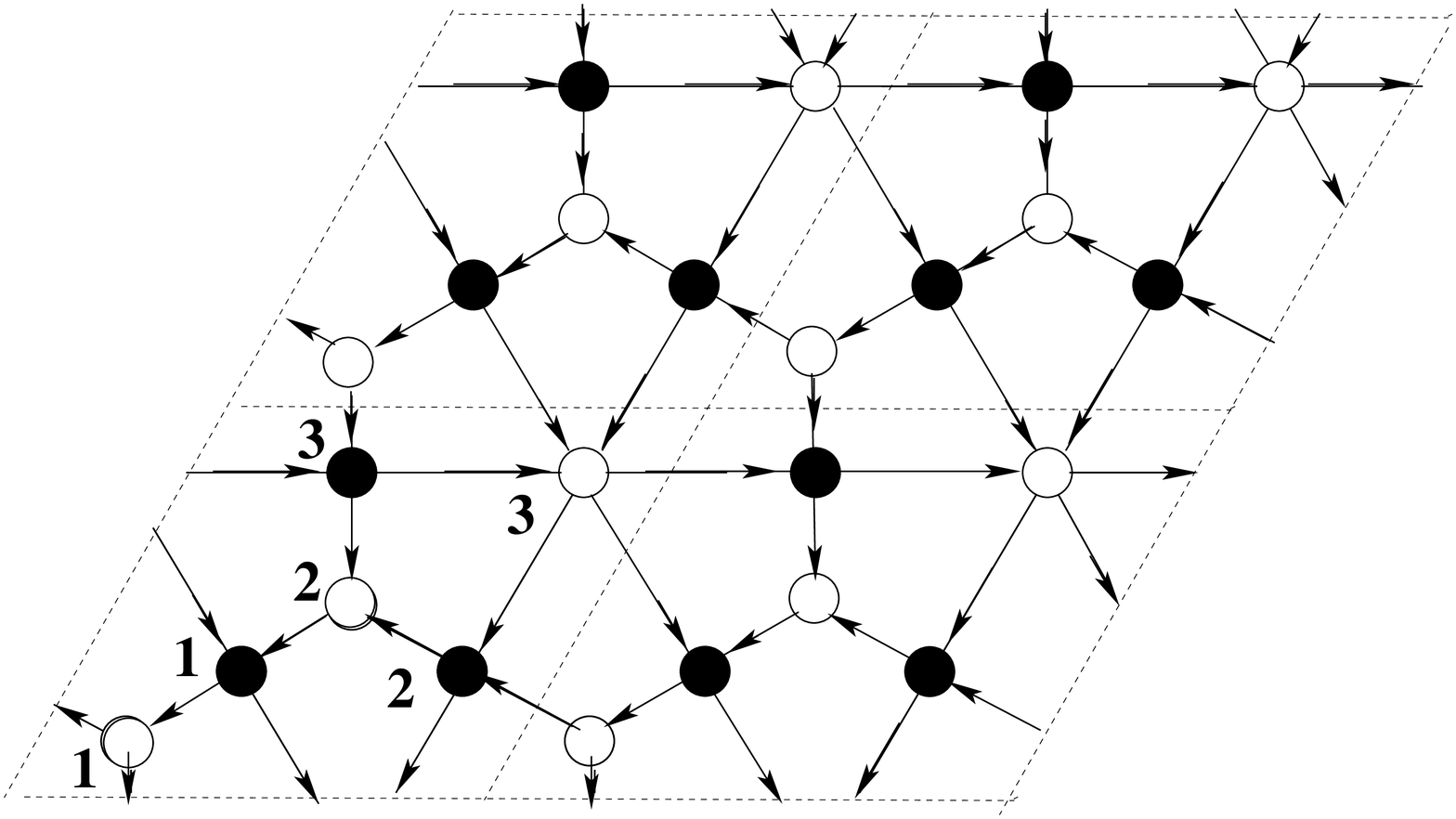}\qquad\qquad\qquad\qquad
\includegraphics[height=30mm]{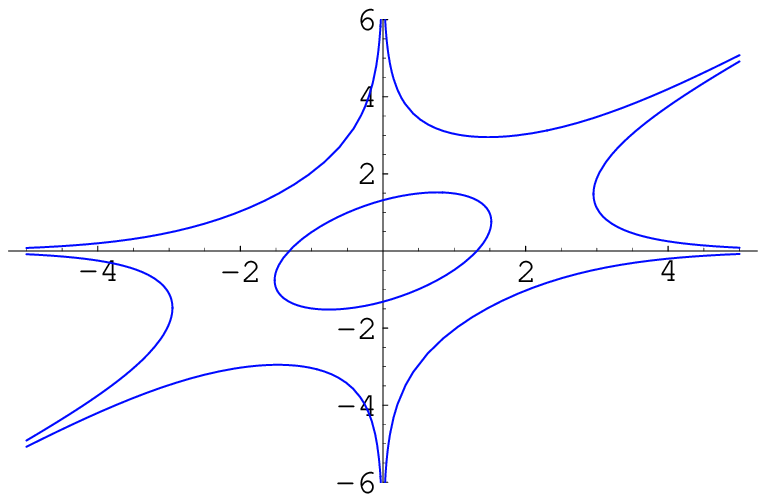}
 \caption{\footnotesize\label{fig:single_oval} Four copies of a Kasteleyn oriented basic tile for the  tunable single oval model and its 
amoeba with $A_{\pm}=B_{\pm}=C_{\pm}=1$ and $D=10$ in 
Eq. (\ref{1oval_polynomial})..}
\end{figure}

For a more complicated tiling (c.f. Fig. \ref{fig:single_oval}) the
analogue of (\ref{spectral_A}) will result in a Fourier transformed
Kasteleyn matrix whose off diagonal block is of size $d$ where $d$ is
the number of vertices of the same colour in the fundamental domain.
The determinant of this matrix will again be a polynomial in $z$ and
$w$ which will determine the phase diagram for the corresponding
bipartite dimer model.  On the amoeba of this polynomial $K$ describes
one massless Dirac Fermion and $d-1$ massive ones, with the massive
modes decoupling from the problem in the thermodynamic limit.

In the thermodynamic limit the logarithm of the bulk partition function per dimer,
$W=\frac{\ln Z}{NM}$, is found to be  
\begin{equation}
W=\int_{0}^{2\pi}\frac{d\theta}{2\pi} \int_{0}^{2\pi}\frac{d\phi}{2\pi} \ln(c-a {\rm e}^{-i\theta}-b{\rm e}^{i\phi})\ .
\end{equation}
If one of the weights is larger than the sum of the other
two, then $W$ is the logarithm of that weight. There are three such regions, 
called frozen regions, where 
one has $W=\ln a$, $W=\ln b$ and $W=\ln c$ \cite{Kasteleyn:1963}. 
The region where none of the weights is larger than the sum of the other two is 
referred to as the amoeba of the spectral curve $p^{Hex}(z,w)$ and is shown in 
Fig. \ref{fig:hex_simple}. On this 
curve there are exactly two pairs of angles $(\Theta,\Phi)$ and $(-\Theta,-\Phi)$ 
at which $p^{Hex}=0$ where, 
\begin{eqnarray}
&&\sin(\Theta)=\frac{b}{2r}\ ,\quad \sin(\Phi)=\frac{a}{2r}\quad {\rm and}\\
&&\kern -24pt r=\frac{a b c}{\sqrt{(a+b+c)(-c+a+b)(c-a+b)(c+a-b)\, .}}
\end{eqnarray}
Here, $r$ is the radius of the circumcircle of the triangle 
with sides $a$, $b$ and $c$ and opposite angles $\Phi$, $\Theta$ 
and $\pi-\Theta-\Phi$ respectively. The radius $r$ diverges, 
while $\Theta$ and $\Phi$ become zero or $\pi$, 
on the boundaries of the amoeba. 

A little effort shows that $W(a,b,c)$ can be expressed in the form
\begin{equation}
W(a,b,c)=\ln c+\frac{\Theta}{\pi}\ln(b/c)
+\frac{1}{2\pi i}({\it li}_2(\frac{a}{c}{\rm e}^{i\Theta})
-{\it li}_2(\frac{a}{c}{\rm e}^{-i\Theta}))
\label{W_dilog}
\end{equation} 
where it was assumed that $c\ge b\ge a$. 
Using the Lobachevsky function \footnote{$L(z)=\frac{1}{2}{\it Cl}_2(2z)$, where
$Cl_2(x)$ is Claussen's function.} 
\begin{equation}
L(z)=-\int_0^z\ln\left\vert2\sin(t)\right\vert dt
\end{equation}
$W$ can be written in the more symmetrical form 
\begin{equation}
W(a,b,c)=p_a\ln a+p_b \ln b+ p_c\ln c+ent(a,b,c)
\end{equation}
where $ent(a,b,c)$ is the entropy per dimer and is given by
\begin{equation}
 ent(a,b,c)=\frac{1}{\pi}\left(L(\pi p_a)+L(\pi p_b)+L(\pi p_c)\right)
\end{equation}
with $p_a+p_b+p_c=1$ and $p_a=\Phi/\pi$ and $p_b=\Theta/\pi$. 

The quantities $p_a$, $p_b$ and 
$p_c$ are
the probabilities that the bonds with activity $a$, $b$ and $c$ respectively, belong to a 
randomly chosen dimer configuration \cite{Cohn:2001}. 

As the edge of the amoeba associated with
say, $c$, is approached $p_c\rightarrow 1$. Curiously $ent(a,b,c)$ 
is the Bloch-Wigner dilogarithm $D(z)$ where 
$z$ can be chosen to be the ratio of any two activities (such that $\vert z\vert\leq1$)
times the exponential of $i$ times the angle between them. This latter quantity has 
$K$-theoretic significance \cite{Nahm:1992sx}.

The internal energy
$U=-\frac{\partial W}{\partial\beta}$ is continuous throughout the phase diagram 
(see \cite{Cohn:2001}) while the specific heat
$C=\beta^2\frac{\partial^2 W}{\partial\beta^2}$ diverges at the transition \cite{Nagle:1989}. 

Explicitly, for $a=b={\rm e}^{-\beta}$ and 
$c=1$, for $\beta > \ln2$ we have $W=U=C=0$, and for $\beta \leq \ln2$ we have
\begin{eqnarray}
&&W=\frac{2}{\pi}\int_\beta^{\ln2}\cos^{-1}(\frac{{\rm e}^{y}}{2})dy ,\\
&&U=\frac{2}{\pi}\cos^{-1}(\frac{{\rm e}^{\beta}}{2})
\quad {\rm and} 
\quad C=\frac{\beta^2}{\pi}\frac{{\rm e}^\beta}{\sqrt{1-\frac{{\rm e}^{2\beta}}{{4}}}} .
\end{eqnarray}
Expanding around $\beta_c=\ln2$, with $\ln2-\beta\ge0$ we have
\begin{eqnarray}
W(\beta)\simeq\frac{4\sqrt{2}}{3\pi}{\left(\ln 2-\beta\right)}^{\frac{3}{2}}.
\end{eqnarray}
In sum: The transition is continuous with no latent heat. 
The specific heat is zero in the low temperature frozen 
phase; there is a phase transition at $\beta=\ln2$,
and the specific heat diverges with critical exponent $\alpha=\frac{1}{2}$
as the transition is approached from the high temperature side.

On the torus corresponding to the $N\times M$ covering by the
fundamental tile the partition function is given by
\begin{eqnarray}
Z=&&
{(-)}^{NM(NM-1)/2}
\frac{1}{2}\sum_{u,v=0}^{1/2}{\rm e}^{2\pi i(\frac{1}{2}+u+v+2uv)} 
{\rm Pfaff}K_{u,v}\nonumber\\
&&=\frac{1}{2}\sum_{u,v=0}^{1/2}{\rm e}^{2\pi i(\frac{1}{2}+u+v+2uv)}{\rm Det} A_{u,v}
\end{eqnarray}
with the four terms corresponding to the four discrete spin structures on the torus.
For the tiling in Fig. \ref{fig:hex_simple}
\begin{eqnarray}
{\rm Det} A_{u,v}&&=\prod_{n=0}^{N-1}\prod_{m=0}^{M-1}
p^{Hex}({\rm e}^{-\frac{2\pi i(n+u)}{N}},{\rm e}^{\frac{2\pi i(m+v)}{M}})\nonumber\\
&&=\prod_{m=0}^{M-1}\left((c-b{\rm e}^{\frac{2\pi i (m+v)}{M}})^N
-a^N {\rm e}^{2\pi i u}\right)
\label{Det_uv}\ .
\end{eqnarray}
The polynomial $p^{Hex}$ has two zeros: One
at $z={\rm e}^{i\Theta},w={\rm e}^{i\Phi}$ the other at $z={\rm e}^{-i\Theta},w={\rm e}^{-i\Phi}$.

Consider, for simplicity, the case when the weights $a$, $b$, and $c$ 
are such that the angles $\Theta$ and $\Phi$ are rational multiples 
of $2\pi$, and in particular 
such that $\Theta=-\frac{2\pi n_0}{N}$ and $\Phi=\frac{2\pi m_0}{M}$, 
for integral $n_0$ and $m_0$. Then shifting $n$ and $m$ appropriately and 
expanding around the zero at $(\Theta,\Phi)$ we have
\begin{equation}
p^{Hex}\simeq-\frac{2\pi i a}{N} {\rm e}^{-i\Theta}(n+u+\tau (m+v))
\end{equation}
where
\begin{equation}
\tau=\frac{N b}{M a}{\rm e}^{i(\Theta+\Phi)} \ .
\end{equation}
In the scaling limit, $N,M\rightarrow\infty$, with the activities and 
$\xi=\frac{N}{M}$ held fixed,  
$\tau$ remains constant and becomes the modular parameter of the continuum limit torus.
The second zero at $(-\Theta,-\Phi)$ simply changes $\tau$ to
$\overline{\tau}$.

Then for large $M$ and $N$ we have 
\begin{eqnarray}
\kern-86pt &&\lim_{N,M\rightarrow\infty}\frac{Z(N,M)}{{\rm e}^{NM W(a,b,c)}}=
Z_{Dirac}(\tau,\theta,\phi)\nonumber\\
&&\qquad\qquad\qquad\qquad=\frac{1}{2}\sum_{u,v=0}^{1/2}
\left\vert\frac{\theta[{}^{\theta+u}_{\phi+v}](0\vert\tau)}{\eta(\tau)}\right\vert
\label{Z_Dirac}
\end{eqnarray}
Where $Z_{Dirac}(\tau,\theta,\phi)$ is the partition function for a Dirac Fermion 
propagating on the continuum torus with modular parameter $\tau$ in the presence
of a gauge potential with zero field strength, i.e. a flat connection, 
but with holonomies ${\rm e}^{2\pi i \theta}$ and
${\rm e}^{2\pi i \phi}$ round the cycles of the torus. 

The two zeros of $p^{Hex}(z,w)$ guarantee that we are dealing with a 
continuum limit for a Dirac Fermion, and in the current tiling the Fermion doubling phenomenon 
is minimized since there are only two vertices in the fundamental tile.
A rectangular toroidal domain arises when $c^2=a^2+b^2$.
which corresponds to $\tau_0=0$ and $\tau_1=\xi\frac{b}{a}$.

For infinite temperature, where $a=b=c=1$ one can see that,
if $N$ and $M$ are positive integral multiples of $6$ 
then ${\rm Det} A_{0,0}=0$ in (\ref{Det_uv}) but nonzero otherwise. 
A similar structure can be seen for other values of $u$ and $v$. 
A careful study of the limiting 
form of (\ref{Det_uv}) therefore requires that the limit be taken 
with the ${\rm mod}~ 6$ periodicity in $N$ and $M$ made explicit.
Setting $N=p~ {\rm mod}~ 6$ and $M=q~ {\rm mod} ~6$
we obtain (\ref{Z_Dirac}) with the modular parameter
$\tau=\xi{\rm e}^{\frac{2\pi i}{3}}$, 
$\theta=\frac{1}{2}-\frac{q}{6}$
and $\phi=\frac{1}{2}+\frac{p}{6}$.

As the temperature is decreased the angle between the periods of the torus,
given by the phase of $\tau$,
reduces continuously reaching zero when the boundary of the amoeba is
reached. For the special case $a/c=b/c={\rm e}^\beta$ 
the modular parameter takes the form
\begin{equation}
\tau\sim\xi\left(1-4(\ln2-\beta)+i 2\sqrt{2}\sqrt{\ln2-\beta}\right)+\dots\ . 
\end{equation}
We see that, at the critical temperature $\beta_c=\ln2$, 
where $\tau_1=0$ while $\tau_0=\xi$, the torus degenerates. It 
effectively reduces to a line. This is a singular limit of the geometry
and is quite distinct from the cylinder limit where for a torus with fixed angle
both $\tau_0$ and $\tau_1$ go to zero.
This phenomenon, of a {\it singular limit as the boundary 
of the amoeba is approached}, is generic.

Note: For $\beta=\frac{\ln 3}{2}$ we have $\tau(\frac{\ln 3}{2})=\frac{1}{2}+i\frac{\sqrt{3}}{2}
=-\frac{1}{\tau(0)}$. With zero holonomies, 
the finite size corrections are modular invariant,
and so unchanged under the replacement $\tau\rightarrow-\frac{1}{\tau}$, however, 
the bulk term is different at these two temperatures. 
 
Note also, that special care is again needed for $\Theta$ and $\Phi$ rational multiples of $2\pi$. 
E.g. when $a=b$ and $\Theta=\frac{2\pi p}{q}$, with $p$ and $q$ relatively prime,
so that $\beta=\ln(2\cos(\Theta))$, then the lattice sizes $M$ and $N$ must be considered 
modulo $q$ and one encounters
a rich structure of flat connections contributing to the finite size corrections.

The dimer model is also closely related to the six vertex model and 
in fact on the free-Fermion line the six vertex model reduces to 
a dimer model. 
In the disordered phase the zero field six-vertex model is critical \cite{Baxter:book}. As the parameters
of the six-vertex model change there is a phase transition, which generically has a latent heat,
to a ferroelectric frozen phase \cite{Lieb:1967bh}. 
But when the phase boundary is crossed from the disordered phase
the specific heat diverges, again with exponent $\alpha=1/2$. 

We conjecture that, for the six-vertex model on a torus, as the
ferroelectric phase is approached from the disordered phase, the
modular parameter, $\tau$, decreases to a critical but non-zero value,
$\tau_c$, and a topological transition occurs.  The special case of the 
dimer model is consistent with this and has $\tau_c=0$ consistent 
with the free-Fermion line.

It was found in \cite{P.Zinn-Justin:2000} 
(see also \cite{Korepin:2000})
that the six-vertex model with domain wall boundary conditions could
be rewritten as a matrix model.  The spectrum of eigenvalues of this
matrix model changes from continuous to discrete as one crosses the
critical line between the disordered and ferroelectric phases.  
Though this matrix realization of the six vertex model 
has not been extended to toroidal
boundary conditions it gives some idea of what is happening 
at a microscopic level as the transition is crossed.

The situation is again reminiscent of what happens in the fuzzy sphere model
\cite{DelgadilloBlando:2007vx}.
However, the high temperature phase is
geometrical with a continuous spectrum in the six-vertex and dimer
models, while the geometrical phase in \cite{DelgadilloBlando:2007vx} 
has discrete spectrum and occurs at low temperatures.
The common feature is the asymmetruc structure and 
characteristic divergence of the specific heat in all of these models.
We expect that this is generic to such topological transitions.

Finally, we observe that the full generality of features of the 
bipartite class of dimer models can be 
accessed in the model whose fundamental tile amd amoeba are shown in 
Fig \ref{fig:single_oval}. 
This figure shows a tiling of the torus whose phase diagram or amoeba 
contains a single compact oval. 

The corresponding spectral polynomial is in general
\begin{equation}
p^{Oval}(z,w)=D-\frac{A_{-}}{z}-A_{+} z-\frac{B_{-}}{w}-B_{+} w-C_{-}\frac{z}{w}-C_{+}\frac{w}{z}
\label{1oval_polynomial}
\end{equation}
where the seven parameters, $A_{\pm}$, $B_{\pm}$ etc.,  are all positive 
and determined by the twelve link activities and can be parametrized as  
$D=2A\cosh(t_x)+2B\cosh(t_y)+2C\cosh(t_{xy})$, 
$A_{\pm}=A{\rm e}^{\pm t_z}$, $B_{\pm}=B{\rm e}^{\pm t_w}$, 
and $C_{\pm}=C{\rm e}^{\pm t_c}$. 
As the oval is crossed the continuum partition function 
describes a massive Dirac Fermion and both the Dirac mass and modular parameter are
tunable by adjusting the activities \cite{Nash_and_OConnor:future}.

\paragraph{Acknowledgements}
This work was partly supported by EU-NCG Marie Curie Network No. MTRN-CT-2006-031962.

\end{document}